\RequirePackage{fix-cm}
\documentclass[smallextended]{svjour3}
\smartqed

\usepackage[T1]{fontenc}
\usepackage{lmodern}
\usepackage{microtype}
\usepackage{xurl}
\usepackage{graphicx}
\usepackage{amsmath,amssymb}
\usepackage{booktabs}
\usepackage{tabularx}
\usepackage{xcolor}
\usepackage{tikz}
\usetikzlibrary{arrows.meta,positioning}
\usepackage[numbers,square,sort&compress]{natbib}
\usepackage[hidelinks]{hyperref}

\journalname{Empirical Software Engineering}

\newcommand{\finding}[1]{\par\medskip\noindent\fbox{\parbox{0.96\textwidth}{\small #1}}\par\medskip}

\begin{document}

\title{Does the Agent's Map Match the Territory?
\subtitle{Architectural Claims in AI Agent Context Files and Their Fit with the Code}}

\author{Bachan Ghimire \and Daniela Damian}

\institute{B. Ghimire \at
  University of Victoria, Victoria, BC, Canada \\
  \email{bachan48@uvic.ca}
  \and
  D. Damian \at
  University of Victoria, Victoria, BC, Canada \\
  \email{danielad@uvic.ca}
}

\date{Received: date / Accepted: date}

\maketitle

\begin{abstract}
AI coding agents start every session without memory of how a system is built. Developers fill this gap with \emph{context files}, such as \texttt{CLAUDE.md} and \texttt{AGENTS.md}. Agents treat these files as true, so it matters whether they match the code. We study two kinds of architectural claims that can be checked against a repository: \emph{path claims}, which name folders and files, and \emph{dependency rules}, which say what may import what. We check 41,090 path claims and 184 dependency rules from 3,602 context files in 3,346 open-source projects against the code and its history. About 5\% of path claims point to paths we cannot find, but we estimate that only 2.2\% are real errors, similar to READMEs (3.5\%). When code moves, 61\% of claims that became stale are fixed, most in the same commit. Dependency rules behave differently: 33 of 184 (18\%) were confirmed broken on the day they were written, usually by older code, and nearly all stayed broken. Context files are good maps but less reliable rulebooks: their rules often describe the architecture a team wants, not the one it has. In an exploratory comparison of 7,480 commits, new rule breaks were rare. Agent-assisted commits broke rules about 1.8 times as often as other commits (0.29\% vs.\ 0.16\%), but with only 14 breaks this is not significant: an early signal, not evidence of a difference or of its absence. We used an LLM with human review at each step, and discuss what this offers and risks for mining studies.
\end{abstract}

\section{Introduction}
\label{sec:intro}

Artificial Intelligence (AI) coding agents such as Claude Code, Codex, Cursor and Copilot now edit repositories with little human help. By early 2026, an estimated 22\% to 29\% of active GitHub projects showed traces of agent use \citep{robbes2026agenticmuch}. Developers no longer ask only for the next line of code. They hand an agent a whole task, and the agent plans the work, edits files and opens a pull request \citep{watanabe2025agentic}. Software development is becoming a form of human-AI teamwork, in which people and agents share one codebase \citep{oneill2020hat,barke2023grounded}.

\textit{Teamwork needs a shared picture of the system.}
Teams coordinate well when members hold overlapping knowledge of the task and of each other. Research calls this a \emph{shared mental model} \citep{cannonbowers1993smm,mohammed2010metaphor}. The same holds for teams of people and automated agents. To be a good team player, an agent must keep \emph{common ground} with the people it works with, so that each side can predict and direct the other \citep{klein2004teamplayer,oneill2020hat}. In software teams, a central part of this shared knowledge is the architecture: how the system is split into parts, and which parts may depend on which. This knowledge helps teams coordinate \citep{espinosa2007team}. Much of it, however, is never written down. Naur~\citep{naur1985theory} argued that it lives in developers' heads. Human newcomers build it over weeks, by reading code and talking to colleagues \citep{kazman2016hdfs}. A coding agent cannot do this. It starts each session without memory of the project. Whatever it should know about the architecture, beyond what it reads in the code, must be handed to it.

\textit{Context files are one place where this hand-over happens.}
Developers write plain-text instruction files that the agent loads at the start of every session. We call them \emph{context files}. The most common are \texttt{CLAUDE.md}, \texttt{AGENTS.md} and \texttt{.github/copilot-instructions.md} \citep{galster2026configuring}. In large part, they are architecture documentation. Architecture appears in 68.1\% of 2,303 context files \citep{chatlatanagulchai2025agentreadmes} and in 72.6\% of 328 Claude Code files \citep{santos2026decoding}. A typical file lists the main folders and what each one holds. It may also state rules such as ``the GUI must not import \texttt{cmk/base}''. 

Figure~\ref{fig:example} shows the kind of text we study. The first part is a \emph{map} of where things are. It is made of \emph{path claims}. The second part is a \emph{rulebook}. It is made of \emph{dependency rules}, which say which parts may import which. In architecture terms, these parts of a context file form an \emph{as-described} architecture \citep{perry1992foundations}, written for a machine reader.

\begin{figure}[b]
\centering
\small
\fbox{\parbox{0.85\textwidth}{\ttfamily
\#\# Project structure\\
- `src/server/routes/` -- HTTP handlers\\
- `src/server/db/` -- database access (Prisma)\\
- `src/client/` -- React front end\\[4pt]
\#\# Architecture rules\\
- Never import from `src/server/` in `src/client/`.\\
- Only `src/server/db/` may import `@prisma/client`.\\
- Do not import `moment`; we are replacing it.
}}
\caption{An illustrative context-file excerpt. The top part holds path claims (the map). The bottom part holds dependency rules (the rulebook).}
\label{fig:example}
\end{figure}

\textit{Wrong architectural knowledge can go unnoticed.}
Human-AI teamwork breaks down when the two sides work from different pictures of the situation \citep{klein2004teamplayer}. Guidelines for human-AI interaction therefore ask that people can see and correct what an AI system believes \citep{amershi2019guidelines}. Context files make this risk concrete. A person who reads a README and finds a missing folder will notice and ask someone. An agent is built to treat the file as true. It re-reads the file every session and may act on it without asking. If the file names folders that no longer exist, the agent may put code in the wrong place. If the file states a dependency rule that the code already breaks, the agent gets mixed signals: the text says one thing, and the code it learns from shows another. In both cases, the error can go unnoticed. This is not only a theoretical concern. Agents already make architectural decisions that nobody reviews \citep{konrad2026architecture}, and context files do change agent behaviour and cost \citep{lulla2026agentsmd,gloaguen2026agentsmd}. If the written knowledge is wrong, human-AI collaboration may speed up architecture erosion instead of preventing it. Whether context files match the code is therefore a question of architecture conformance and evolution. Answering it is a first step toward better ways of sharing architecture with agents.

\textit{What we know so far.}
It is well known that described and implemented architectures drift apart \citep{perry1992foundations,desilva2012erosion,li2022erosion}. Researchers have tools to measure the gap, such as reflexion models and dependency rules \citep{murphy2001reflexion,knodel2007comparison,sangal2005dsm}. Recent work has started to ask whether context files go stale, a problem called \emph{context rot} \citep{treude2026contextrot}. That study checked whether names of code elements in context files still exist. It left architecture-level consistency as future work. Other work checks whether LLMs can detect violations of architecture decision records \citep{su2026adrviolations}, or turns context-file rules into executable checks \citep{sharma2026contextcov}. To our knowledge, no study has measured at scale whether the structural and dependency claims written for agents match the code, or how these claims change over a project's history.

\textit{Scope and research questions.}
We study two kinds of \emph{architectural claims} that can be checked against a repository:
\begin{itemize}
\item A \emph{path claim} mentions a folder or file in the repository. Example: ``API handlers live in \texttt{src/server/routes/}.''
\item A \emph{dependency rule} says that one part of the system must not (or may only) import another part or a library. Example: ``code in \texttt{pgqueuer/core/} must not import from \texttt{pgqueuer/adapters/}.''
\end{itemize}
Both kinds of claims can be checked against the code, and allowed dependencies are the classic object of architecture conformance checking (Section~\ref{sec:construct}). 
Code changes arrive as commits, and today some of these commits are made with the help of a coding agent. This lets us ask whether such commits break the rules more often. An \emph{agent-assisted commit} is a commit with a visible trace of a coding agent, such as a \texttt{Co-authored-by: Claude} line. We ask three questions:

\begin{description}
\item[\textbf{RQ1:}] How accurate are the path claims and dependency rules in context files, and how do path claims compare with READMEs in the same repositories?
\item[\textbf{RQ2:}] Do path claims in context files stay accurate as the code changes?
\item[\textbf{RQ3:}] What happens to dependency rules after they are written, and do agent-assisted commits break them more often than other commits?
\end{description}

RQ1 asks whether the claims are right at one point in time. READMEs are the most common description of a project written for people, so they give a baseline for path claims. RQ2 asks whether context files keep up with the code. Classic documentation often lags behind code \citep{aghajani2019issues,wen2019comments}, and many context files are written once and rarely touched \citep{mohsenimofidi2026context,denisovblanch2026ramp}. RQ2 also separates claims that were wrong from the start from claims that became stale later. RQ3 follows dependency rules over time: whether they were broken when written, whether they stayed broken, and whether later commits broke or repaired them. Its second part, the comparison of agent-assisted and other commits, is exploratory. It is a first, observational look at whether rules written for agents are broken in commits that agents helped make. Benchmarks show that agents often miss design constraints that tests do not catch \citep{yu2026passrate,orlanski2026slopcodebench}, and that agents follow instructions less well as a session goes on \citep{mcmillan2026adherence}. We do not yet know how this plays out in real projects.

\textit{How we answer the questions.}
We start from a public dataset of 36,710 engineered GitHub projects \citep{galster2026dataset}, meaning real software, not tutorials or toy repositories. We take every project, in any language, that has a project-level context file. We analyze the 3,346 projects whose files make at least one path claim. For RQ1, we check path claims against the file tree and review samples by hand to separate real errors from extraction noise. For RQ2, we replay the history of each context file and README, commit by commit. For RQ3, we turn each dependency rule into a check that a program can run. An \emph{import checker} reads the import statements in each source file and tests every later commit that touched the code the rule covers. We find agent-assisted commits from traces in the commit message, the author account and the pull request.

\textit{Contributions.}
This paper makes four contributions:
\begin{enumerate}
\item A large-scale measurement of path claims and dependency rules in agent context files against the code: 41,090 path claims and 184 dependency rules in 3,346 projects and ten languages, with READMEs as a baseline.
\item A commit-level history analysis that separates claims that were wrong when written from claims that became stale later, and measures how fast claims go stale and get fixed.
\item A history of dependency rules after they are written, with an exploratory comparison of rule breaks in agent-assisted and other commits in real repositories. We detect agent-assisted commits from commit, account and pull-request traces.
\item A transparent account of how we used LLMs in a mining study, how we checked their output by human review, and what we learned about the benefits and risks of this practice.
\end{enumerate}

\textit{Structure of the paper.}
Section~\ref{sec:related} covers background and related work. Section~\ref{sec:method} describes the study design, including our use of LLMs and our human review. Section~\ref{sec:results} answers the research questions. Section~\ref{sec:discussion} discusses the findings, our experience with LLMs as research tools, and implications. Section~\ref{sec:threats} covers threats to validity, Section~\ref{sec:future} future work, and Section~\ref{sec:conclusion} concludes.

\section{Background and Related Work}
\label{sec:related}

\subsection{Agent context files}

Coding agents can be configured in several ways at the repository level. Plain Markdown context files are the most common, and often the only, mechanism \citep{galster2026configuring}. \texttt{AGENTS.md} is becoming a shared format across tools \citep{galster2026configuring,mohsenimofidi2026context}.

A fast-growing body of work describes what these files contain. Chatlatanagulchai et al.~\citep{chatlatanagulchai2025manifests} studied 253 \texttt{CLAUDE.md} files, and later 2,303 context files from 1,925 repositories \citep{chatlatanagulchai2025agentreadmes}. Build and test commands, implementation details and architecture were the most common content. Security and performance were rare. Santos et al.~\citep{santos2026decoding} found architecture sections in 72.6\% of 328 Claude Code files. Mohsenimofidi et al.~\citep{mohsenimofidi2026context} studied \texttt{AGENTS.md} files in 466 projects. They found no settled structure and five ways of giving instructions to agents, including \emph{prohibitive} ones. Other work catalogues \emph{configuration smells}: flaws, redundancies and unhelpful practices in these files that can lower code or agent quality \citep{santos2026smells}. Files keep growing because old instructions are rarely deleted \citep{chakrabarti2026claudemd}, and they rarely say how the system is expected to evolve \citep{chondamrongkul2026evolution}.

Studies of other rule files for agents point the same way. A census of 22,854 popular GitHub repositories found agent instructions in 26.3\% of them. Most directives (74\%) give no reason, and four in five instruction sets were edited after they were created \citep{zarzycki2026census}. More than 12,000 \texttt{.cursorrules} files were dominated by guidance on code quality, project structure and maintainability \citep{sun2026cursorrules}. Cai et al.~\citep{cai2026rules} mined 7,310 rules from AI IDE projects and surveyed 99 practitioners. Practitioners rated architectural constraints as highly important, yet the rule files held mostly low-level workflow and formatting rules. Developers said they changed rules mainly to correct AI errors, usually by adding new negative constraints. For 160 rule updates, Cai et al. also checked the affected code and found that it followed the rules more often after an update. These studies describe what the rules say, how they change and, in one case, whether code follows recent edits. None checks architectural rules against the code at scale, or from the day each rule was written.

A second line of work asks whether context files help, with mixed results. In one study, \texttt{AGENTS.md} files reduced agent runtime and output tokens \citep{lulla2026agentsmd}. Other controlled studies found little or no gain in task success, and higher cost \citep{gloaguen2026agentsmd,khatri2026contextfiles}. Gloaguen et al.~\citep{gloaguen2026agentsmd} report that agents follow explicit instructions well, but that repository overviews do not help. Zhang et al.~\citep{zhang2026guardrails} found that the most helpful rules are negative constraints, such as ``do not refactor unrelated code''. Dependency rules are exactly such constraints. McMillan~\citep{mcmillan2026adherence} found that file structure does not change how well agents follow instructions, but that adherence drops as a session goes on.

Closest to our work, Treude and Baltes~\citep{treude2026contextrot} named the problem of stale context \emph{context rot}. They applied an existing checker for outdated code-element references \citep{tan2024outdated} to 356 repositories and found stale references in 23.0\% of them. Their study works at the level of identifiers. It leaves architecture-level consistency as an open question. Sharma~\citep{sharma2026contextcov} compiles constraints from instruction files into static checks and enforces them on a benchmark. Neither study measures how well structural and dependency claims fit real code over time.

\subsection{Coding agents and the quality of their changes}

Agent traces in commits and pull requests make mining studies of agent activity possible \citep{robbes2026perils}. Pull requests opened by agents are often merged, but they still need human fixes, often to meet project-specific standards \citep{watanabe2025agentic,li2025aidev}. In long iterative tasks, code written by agents erodes structurally faster than code written by people \citep{orlanski2026slopcodebench}. Passing tests also says little about whether a patch respects design constraints \citep{yu2026passrate}. At the architecture level, Larsen and Moghaddam~\citep{larsen2026mining} found no change in the number of architectural smells after Java projects adopted agents. Konrad et al.~\citep{konrad2026architecture} argue that agents now make architectural decisions that nobody reviews.

Controlled studies show similar problems. When a task adds more structural constraints, such as required layers or data-access patterns, the share of tests that agents pass drops by 27 points on average \citep{dente2026constraint}. More capable models tend to write more bloated and coupled code, and neither functional correctness nor detailed prompting prevented this \citep{zhu2026smells}. In contrast, giving an LLM an architecture document improved how well its code conformed to the intended design \citep{cervantes2026archdocs}. Written architectural guidance can help, then, but only if it is right.

\subsection{LLMs and software architecture}

LLMs have been used to generate design decisions \citep{dhar2024adr}, to answer questions about a system's architecture \citep{soliman2025gptak}, and to support many other architecture tasks \citep{schmid2025salm,bucaioni2025aisa}. Conformance checking is still little studied \citep{schmid2025salm}. Su et al.~\citep{su2026adrviolations} showed that LLMs can detect violations of architecture decision records well when the decision is visible in the code. We borrow their idea of using an LLM for extraction with human review. Our goal is different, though. We measure how well agent-facing claims fit the code, not how well an LLM checks them. Hou et al.~\citep{hou2024llm4se} survey LLMs in software engineering more broadly.

\subsection{Architecture erosion and documentation drift}

Perry and Wolf~\citep{perry1992foundations} described \emph{drift} and \emph{erosion} as the growing gap between the intended and the built architecture. Reflexion models compare a high-level model with the source code and report where they agree and differ \citep{murphy2001reflexion}. Dependency rules and design structure matrices state which parts may use which \citep{sangal2005dsm,knodel2007comparison,passos2010conformance}. Surveys summarize the causes of erosion and ways to control it \citep{desilva2012erosion,li2022erosion}. An industrial study found that fewer than half of detected violations were removed four months later. Architects often changed the model rather than the code \citep{buckley2015reflexion}.

For documentation in general, being out of date is a top problem \citep{aghajani2019issues,aghajani2020practitioners}. Comments lag behind code \citep{wen2019comments}, and most projects contain outdated code references at some point \citep{tan2024outdated}. Recent tools check each commit for such errors and propose fixes, for example by comparing a README with the code it describes \citep{baek2026readu}. In open source, architecture documentation is informal, rarely updated and often kept by one person \citep{migliorini2024views}. Decision records are rare \citep{buchgeher2023adr}, and READMEs mix many kinds of content \citep{prana2019readme}. Architecture documentation mainly helps newcomers \citep{kazman2016hdfs}. An agent is a newcomer at the start of every session.

\subsection{Architectural rules in practice}
Architectural rules state which parts of a system may depend on which. In practice, they are often defined but rarely tested \citep{caracciolo2016monitoring}. In a study of 395 developers, architecture documentation was mostly incomplete, informal and outdated, and conformance checks were seldom done, let alone automated \citep{melo2016perceptions}. In three industrial projects, architects also stated kinds of rules that common conformance tools do not cover \citep{schroeder2019discovering}. Conformance tools are underused, and the violations they find are rarely fixed. Where a check runs in continuous integration, however, violations can be caught and fixed before each release \citep{pinto2017archci}. The tools have limits too. A benchmark of ten checkers found that they reported, on average, 77\% of the dependencies in test code, and named ten kinds of dependency that are hard to detect \citep{pruijt2017accuracy}. Our import checker faces the same problem. Other work avoids writing a formal model at all. Maffort et al.~\citep{maffort2016mining} find likely violations from version history with simple heuristics, with a precision of about 54\% to 63\% on three systems. Lucena et al.~\citep{lucena2025gatekeepers} use LLMs to turn design rules discussed in GitHub pull requests and issues into executable design tests. Context files offer a new source of such rules: written by developers, in natural language, next to the code, and for a machine reader. Unlike rules kept by an architect, these rules are read and acted on by an agent in every session, so whether they are true matters more.

When the code does not follow the intended architecture, the gap is often a form of \emph{architectural technical debt}. Such debt comes from large design decisions that later hinder progress, and it is hard to find and costly to remove \citep{verdecchia2021atd}. Some debt-incurring design decisions are taken on purpose \citep{soliman2021dadd}, and debt tends to build up during migrations, such as moves to microservices \citep{detoledo2022migration}. In open source, design decisions often drift from the architect's intent, and the resulting architectural smells raise maintenance effort \citep{le2018decay}. Developers sometimes admit such gaps in code comments. Sierra et al.~\citep{sierra2019satd} found that these self-admitted technical debt comments can point to architectural divergences, although tracing them takes much effort. A rule in a context file that the code still breaks can be read in a similar way: as a written, but unmarked, record of intended architecture and of debt still to be paid.

\subsection{Human-AI collaboration and shared understanding}

Human factors research has studied how people work with automated systems, and more recently with AI agents. An agent is a good teammate when people can predict what it will do and direct it when needed, which requires common ground \citep{klein2004teamplayer}. Reviews of human-autonomy teaming name shared cognition and information sharing among the factors that most affect team outcomes \citep{oneill2020hat}. Experiments link shared mental models between humans and agents to performance and trust \citep{schelble2022letsthink}. Guidelines for human-AI interaction ask that AI systems make clear what they know and let people correct them \citep{amershi2019guidelines}. In programming, developers switch between letting a code model speed them up and using it to explore. Either way, they must judge its output against their own understanding of the code \citep{barke2023grounded}.

For architecture, early work explored people and chatbots designing a system together \citep{ahmad2023humanbot}. Agents still struggle to keep a stable picture of a codebase's architecture \citep{sapunov2026tocs}, and structured architecture descriptions help weaker models most \citep{canedo2026equalizer}. Recent work calls for shared knowledge spaces that link a team's unwritten knowledge with system artifacts \citep{dearstyne2026darkmatter}, and for hybrid intelligence, where people and AI combine their strengths \citep{akata2020hybrid}. A context file is one concrete, versioned example of such a space. We study how well it carries structural and dependency knowledge.

\begin{table}[h]
\centering
\caption{Closest prior work and what this study adds.}
\label{tab:positioning}
\small
\begin{tabularx}{\textwidth}{@{}>{\raggedright\arraybackslash}p{3.3cm}>{\raggedright\arraybackslash}X>{\raggedright\arraybackslash}X@{}}
\toprule
Study & What it measures & What we add\\
\midrule
Content studies \citep{chatlatanagulchai2025agentreadmes,santos2026decoding,mohsenimofidi2026context} & Which topics context files cover & Whether structural claims match the code\\
Effect studies \citep{gloaguen2026agentsmd,lulla2026agentsmd,khatri2026contextfiles} & Agent success and cost with or without files & Whether file content is accurate, in real projects\\
Context rot \citep{treude2026contextrot} & Stale identifiers at one point in time & Folders, files and rules, across history\\
ContextCov \citep{sharma2026contextcov} & Enforcing rules on a benchmark & Whether rules hold in real repositories\\
ADR violations \citep{su2026adrviolations} & How well an LLM spots decision violations & Rules written for agents; agent-assisted vs.\ other commits\\
Agent adoption and architecture \citep{larsen2026mining} & Smell counts after adoption & Breaks of stated rules\\
Rule-file studies \citep{zarzycki2026census,cai2026rules} & What agent rules say, how they change, and compliance after rule updates & Whether architectural rules are true of the code, from the day they are written\\
Conformance checking \citep{maffort2016mining,lucena2025gatekeepers} & Violations of rules from experts, history or discussions & Rules developers wrote for agents, at scale and across history\\
Self-admitted debt \citep{sierra2019satd} & Whether code comments point to architectural divergences & Rules that are broken on the day they are written\\
\bottomrule
\end{tabularx}
\end{table}

\subsection{Positioning}

Table~\ref{tab:positioning} places our study among the closest prior work. Earlier studies describe what context files say, test whether they help agents, or check single rules, names or decisions, often on benchmarks or with expert-written rules. We instead check the folders, files and dependency rules that developers write for agents against the real code, across many projects and their history, and find that many rules are broken from the day they are written.

\section{Study Design}
\label{sec:method}

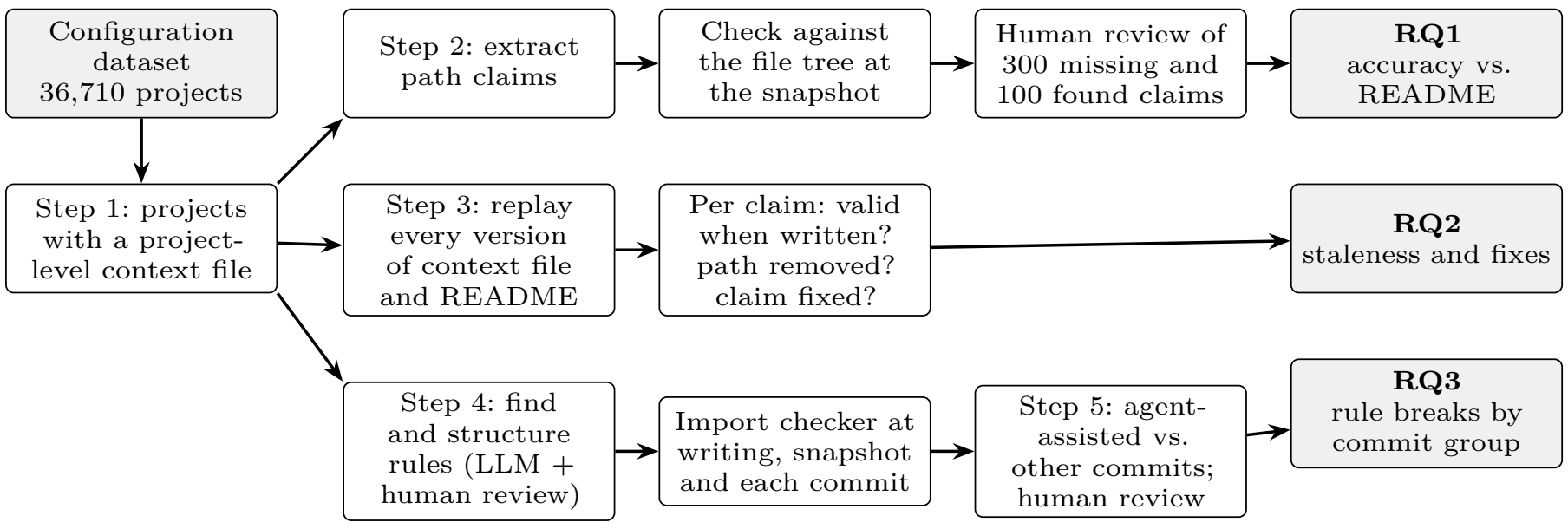
\begin{figure}[h]
\centering
\resizebox{\textwidth}{!}{%
\begin{tikzpicture}[
  node distance=0.35cm and 0.4cm,
  box/.style={draw, rounded corners=2pt, align=center, font=\scriptsize, minimum height=1.0cm, text width=2.3cm, fill=white},
  data/.style={box, fill=gray!12},
  arr/.style={-{Stealth[length=2mm]}, thick}]
\node[data] (ds) {Configuration dataset\\ 36,710 projects};
\node[box, below=0.6cm of ds] (sel) {Step 1: projects with a project-level context file};
\node[box, right=0.6cm of ds] (ext) {Step 2: extract path claims};
\node[box, right=of ext] (chk) {Check against the file tree at the snapshot};
\node[box, right=of chk] (val) {Human review of 300 missing and 100 found claims};
\node[data, right=of val] (rq1) {\textbf{RQ1}\\ accuracy vs.\ README};
\node[box, below=0.6cm of ext] (hist) {Step 3: replay every version of context file and README};
\node[box, right=of hist] (fate) {Per claim: valid when written? path removed? claim fixed?};
\node[data, below=0.6cm of rq1] (rq2) {\textbf{RQ2}\\ staleness and fixes};
\node[box, below=0.6cm of hist] (rules) {Step 4: find and structure rules (LLM + human review)};
\node[box, right=of rules] (imp) {Import checker at writing, snapshot and each commit};
\node[box, right=of imp] (who) {Step 5: agent-assisted vs.\ other commits; human review};
\node[data, below=0.6cm of rq2] (rq3) {\textbf{RQ3}\\ rule breaks by commit group};
\draw[arr] (ds)--(sel);
\draw[arr] (sel.north east) -- (ext.south west);
\draw[arr] (ext)--(chk); \draw[arr] (chk)--(val); \draw[arr] (val)--(rq1);
\draw[arr] (sel)--(hist); \draw[arr] (hist)--(fate); \draw[arr] (fate)--(rq2);
\draw[arr] (sel.south east) -- (rules.north west); \draw[arr] (rules)--(imp); \draw[arr] (imp)--(who); \draw[arr] (who)--(rq3);
\end{tikzpicture}}
\caption{Study overview. Grey boxes are inputs and results.}
\label{fig:pipeline}
\end{figure}

Figure~\ref{fig:pipeline} gives an overview. The study has five steps:
\begin{enumerate}
\item We select projects and their context files.
\item We extract path claims and check them against the code (RQ1).
\item We replay each claim's history (RQ2).
\item We find dependency rules and turn them into import checks (RQ1, RQ3).
\item We compare the commits that break rules (RQ3).
\end{enumerate}

After describing the five steps, we explain where we used an LLM (Section~\ref{sec:llmuse}), which robustness checks we ran (Section~\ref{sec:robust}), and how we reviewed the output of each step by hand (Section~\ref{sec:checks}).

\subsection{What we measure, and why}
\label{sec:construct}

Architecture has many views. We measure two parts of it that context files state often and that can be checked against the code mechanically, once a rule is mapped to folders.

\paragraph{Path claims describe the module structure.} They say which folders and files exist and what each holds. Layout of this kind is among the most common content of context files \citep{chatlatanagulchai2025agentreadmes,santos2026decoding}. A wrong path gives an agent a wrong picture of where code lives. Not every path in a context file is architectural, so we also report claims under structure or architecture headings separately (Table~\ref{tab:breakdown}).

\paragraph{Dependency rules describe allowed dependencies.} They say which parts of the system may use which. Allowed dependencies are the classic object of static architecture conformance checking: reflexion models, dependency rules and design structure matrices all compare stated dependencies with those found in the code \citep{murphy2001reflexion,knodel2007comparison,sangal2005dsm,passos2010conformance}. Conformance checkers find these dependencies mainly through import statements \citep{pruijt2017accuracy}, and so do we.

\paragraph{Our measure is conservative.} We count only static imports in production code. This misses dependencies declared only in build files or loaded at runtime. It can therefore miss breaks. Wrong structured rules or import resolution can also report breaks that are not there, which human review removes (Section~\ref{sec:step4}). Our confirmed counts of broken rules are therefore lower bounds. We check the checker in both directions: flagged breaks for false alarms, and rules that held for missed breaks (Section~\ref{sec:checks}).

\subsection{Snapshot and checkpoints}
\label{sec:time}

We track change commit by commit, not at fixed intervals such as days or weeks. Two terms help here.

The \emph{snapshot} is one fixed point per project: the last commit on the scanned branch before the dataset's scan date (early 2026). RQ1 measures accuracy at the snapshot.

A \emph{checkpoint} is a commit at which we look at a claim during its history. There are up to four per claim: the commit where the claim was written, the commit where it was removed from the file (if any), the last commit that touched its path, and the snapshot. RQ2 compares a claim's state across its checkpoints.

All times are commit timestamps, reported in days. Our observation window runs from two days before a project's first context file was created to the snapshot. The median window is 67 days per file. For RQ3, we look at every commit in the window after a rule was written.

\subsection{Step 1: Projects and context files}
\label{sec:sample}

We use the public dataset of Galster et al.~\citep{galster2026dataset}, collected in early 2026. It starts from licensed, active GitHub repositories with several contributors. An LLM sorted these repositories into \emph{engineered} projects, meaning real software rather than tutorials, homework or personal configuration repositories \citep{munaiah2017curating}. This gave 36,710 engineered projects. The dataset also lists their agent configuration files and 148,519 commits with an AI co-author line. We did not repeat this classification; we use the dataset as published.

We keep every engineered project, in all ten main languages of the dataset. From each project, we take the context files that coding agents load by default:
\begin{itemize}
\item \texttt{AGENTS.md};
\item \texttt{CLAUDE.md}, at the root or in \texttt{.claude/};
\item \texttt{.github/copilot-instructions.md};
\item \texttt{GEMINI.md};
\item \texttt{.cursorrules}.
\end{itemize} We drop files shorter than 20 lines and files that only point to another file. This gives 4,372 files in 3,909 projects. We could clone the history of 3,836 of these projects. Of these, 3,346 projects with 3,602 context files make at least one path claim. They form our sample (Table~\ref{tab:sample}). About a quarter of the files (25\%) were first committed with a visible agent trace. Step~5 explains how we detect such traces.

\begin{table}[t]
\centering
\caption{Sample overview.}
\label{tab:sample}
\small
\begin{tabular}{@{}lr@{}}
\toprule
Characteristic & Value\\
\midrule
Projects / context files & 3,346 / 3,602\\
\quad \texttt{AGENTS.md} / \texttt{CLAUDE.md} / Copilot / other\textsuperscript{a} & 1,463 / 1,307 / 780 / 52\\
Main language: TypeScript / Python / Go & 817 / 637 / 474\\
\quad C\# / Rust / Java / JavaScript & 247 / 233 / 232 / 232\\
\quad C++ / PHP / C & 197 / 192 / 85\\
Stars / contributors / commits (median) & 530 / 36 / 1,843\\
Context file length (median lines) & 142\\
\bottomrule
\multicolumn{2}{@{}l}{\footnotesize \textsuperscript{a}\texttt{GEMINI.md} and \texttt{.cursorrules}. \texttt{CLAUDE.md} includes 78 files in \texttt{.claude/}.}
\end{tabular}
\end{table}

\subsection{Step 2: Path claims (RQ1)}
\label{sec:step2}

\paragraph{Extraction.}
A rule-based extractor finds path claims in two forms.
\begin{itemize}
\item An \emph{inline claim} is a path written as code, such as \texttt{`src/server/routes/`}. We keep such a token if it contains a slash and either ends with a slash, has a common file extension, or starts with a top-level folder such as \texttt{src/}.
\item A \emph{tree claim} comes from a folder diagram drawn with box characters, like the output of the \texttt{tree} command. We rebuild each full path from the indentation.
\end{itemize}
We drop URLs, absolute paths, package names, wildcards and placeholders (such as \texttt{\{name\}}), common build and cache folders, and paths in \texttt{.gitignore}. For each claim, we note whether it sits under a heading about structure or architecture, such as ``Project structure''. This is the case for 63\% of context-file claims. Folder trees account for 26\% of claims.

\paragraph{Checking.}
A claim is \emph{found} if its path exists in the file tree at the snapshot. The path can match from the root, or as the end of a longer path. We call the second option \emph{lenient matching}. It avoids penalizing a file that describes one package of a larger repository. A claim that is not found is \emph{missing}. We run the same extractor and checker on the README at the snapshot.

\paragraph{Estimating real errors.}
The extractor can also pick up example paths, paths in the user's own project, or build outputs. To estimate real errors, we reviewed a random sample of 300 missing claims: 180 from context files and 120 from READMEs. Each claim received one of five labels (Table~\ref{tab:validation} in Section~\ref{sec:rq1}). Only claims that the text presents as an existing part of the repository count as \textsc{Wrong path}. We then estimate the real error rate in three short steps:
\begin{enumerate}
\item The share of real errors among the sampled missing claims is the \emph{precision} of our checker.
\item We compute a 95\% Wilson interval for this precision \citep{wilson1927probable}.
\item We multiply the raw missing rate by the precision, and carry the interval over.
\end{enumerate}
We also reviewed 100 \emph{found} claims, to see whether lenient matching accepts wrong claims.

\subsection{Step 3: Claim history (RQ2)}

For each project, we clone the scanned branch for the observation window. We replay every commit that changed a context file or README, and record each claim's checkpoints (Section~\ref{sec:time}). This gives each claim one of the following outcomes:
\begin{itemize}
\item \emph{Valid when written}: its path existed at the commit that added the claim.
\item \emph{Never valid}: its path did not exist when the claim was written.
\item \emph{Gone stale}: it was valid when written, and its path later disappeared while the claim was still in the file.
\item A stale claim is \emph{fixed} if it was later removed or corrected, and \emph{unfixed} if it was still there at the snapshot.
\end{itemize}
A valid claim may also be deleted from the file for other reasons while its path still exists. We do not check the file tree at every commit in between. So a path that is deleted and later restored does not count as a change.

Claims were followed for different lengths of time. Besides the share of claims that went stale, we therefore report a rate per \emph{claim-month}. One claim followed for one month counts as one claim-month.

\subsection{Step 4: Dependency rules (RQ1, RQ3)}
\label{sec:step4}

\paragraph{Finding rules.}
A dependency rule says that some files must not import something, or that only some files may import it. Finding them takes three stages.
\begin{enumerate}
\item \emph{Keyword filter.} A broad pattern selects candidate sentences from all context files. It looks for restriction words (\emph{never}, \emph{must not}, \emph{only}) near dependency phrases (\emph{import}, \emph{depend on}), for statements about layers, and for names of import-linting tools. This gives 1,645 sentences in 875 projects.
\item \emph{LLM classification and structuring.} An LLM (Claude Opus~5.5) decided whether each sentence is a dependency rule. For each rule, it wrote a structured version. This lists the folders the rule covers, any exempt folders, and the forbidden folders or external packages. The LLM saw the project's folder list, so it could map names such as ``the GUI'' to real paths. It marked unclear mappings as uncheckable.
\item \emph{Human review.} The first author reviewed every structured rule against its sentence and the repository, and corrected or rejected it where needed. This review also marked the eight rules that allow an import only from certain places, and the rules that apply to each sibling folder separately. A second reviewer verified a sample (Section~\ref{sec:checks}).
\end{enumerate}
Of the 1,645 sentences, 357 are dependency rules, and 176 of these can be checked against the code. Some sentences hold several rules, for example ``never import A into B or vice versa''. After splitting such sentences and removing duplicates, we have 184 checkable rules in 115 projects. We leave out rules about import style that do not separate parts of the system, such as ``no wildcard imports''.

\paragraph{Validating the structured rules.}
A wrong structured rule can fail in two ways. It can make the checker report breaks that are not there, or hide breaks that are. We checked both.
\begin{itemize}
\item \emph{False breaks.} We reviewed all 51 rules flagged as broken when written. In 15, the structured form did not match the sentence or missed a stated exception. In 2 more, an import was resolved to the wrong target. We removed these from the confirmed counts.
\item \emph{Hidden breaks.} We reviewed 80 of the 129 rules that held: 40 for missed breaks and 40 more for the fit between sentence and structured form. Two of the 80 had a structured form that did not fully express the sentence.
\end{itemize}
Mapping errors were thus common among rules flagged as broken, where review removed them, and rare among rules that held. Section~\ref{sec:checks} gives the details and the agreement with a second reviewer.

\paragraph{Checking imports.}
Our \emph{import checker} reads the import statements in each source file and tests them against the rules. It handles eleven language families: Python, JavaScript/TypeScript, Go, Java/Kotlin/Scala, C\#, Rust, PHP, Ruby, C/C++, Swift and Dart. It resolves each import to a folder in the repository or to an external package. To do this, it uses relative paths, path aliases, module and package names from build files such as \texttt{go.mod}, and, for Java and C\#, folders whose names match the package.

An import \emph{breaks} a rule if the importing file is covered by the rule and the import points to a forbidden target. We ignore type-only imports. We also ignore test, example, script and vendored files, because rules are usually meant for production code. We run the checker at the commit where each rule was written and at the snapshot.

\subsection{Step 5: Commit analysis (RQ3)}
\label{sec:step5}

\paragraph{Exposed commits.}
After a rule is written, every non-merge commit that changes at least one file covered by the rule could break it. We call these \emph{exposed commits}. A commit \emph{adds a break} if it adds a rule-breaking import. If the same commit removes that import elsewhere, the code was only moved, and it does not count. A commit \emph{repairs} a rule if it removes a rule-breaking import without adding it back. We follow file renames. We skip the oldest commit of each clone, because its diff would show all existing code as new.

\paragraph{Commit groups.}
We split exposed commits into two groups. A commit is \emph{agent-assisted} if it shows a visible trace of a coding agent. All remaining commits, except those of dependency and CI bots, are \emph{other commits}. A co-author line alone misses many agent commits \citep{robbes2026perils,khosravani2026census}. We therefore combine three kinds of traces, based on the published heuristics of Robbes et al.~\citep{robbes2026perils} (Table~\ref{tab:detect}): the commit message, the author or committer account, and the pull request. Pull-request traces add 331 commits, 24\% more than message traces and accounts alone.

\begin{table}[t]
\centering
\caption{Traces used to detect agent-assisted commits.}
\label{tab:detect}
\small
\begin{tabularx}{\textwidth}{@{}l>{\raggedright\arraybackslash}Xr@{}}
\toprule
Trace & What we look for & Commits\\
\midrule
Commit message & In the dataset's list of AI co-authored commits: an agent co-author line, a ``Generated with'' line, or an agent session link & 1,332\\
Agent account & Author or committer is an agent account (e.g., \texttt{copilot-swe-agent}) & +65\\
Pull request & A traced commit in the same squash-merged or merged pull request, or a branch named after an agent & +331\\
\midrule
Total & & 1,728\\
\bottomrule
\end{tabularx}
\end{table}

\paragraph{Comparing the groups.}
New rule breaks are rare, and commits from the same project tend to be alike. We therefore compare the groups within each project and then combine the projects using the Mantel-Haenszel odds ratio \citep{mantel1959statistical}. An odds ratio of 1 means no difference between the groups. This is the only statistical test in RQ3, and we run it on the breaks confirmed by human review.

\subsection{Use of LLMs}
\label{sec:llmuse}

We used one LLM, Claude Opus~5.5 (Anthropic), in three roles. In every role, a person checked the output before it was used.

\paragraph{Role 1: Finding and structuring rules (Step 4).}
The LLM sorted the candidate sentences into rules and non-rules, and wrote the structured form of each rule (Section~\ref{sec:step4}). It did not decide whether a rule holds in the code. The import checker, an ordinary program, did that.

Why use an LLM here at all? A keyword pattern alone works poorly. Our broad filter keeps 1,645 sentences, but only 22\% of them are dependency rules. A stricter pattern, which requires both a restriction word and an import word in the same sentence, keeps 952 sentences. Only 28\% of these are rules, and the pattern misses 29 of the 148 sentences behind our 184 checked rules (20\%). The LLM step keeps the rules without this noise, and the first author then reviewed every structured rule.

\paragraph{Role 2: Proposing review labels (Section~\ref{sec:checks}).}
For each review sample, the LLM proposed a label and a short reason that pointed to evidence, such as a file and line. It did this twice, as two \emph{labellers} (A and B), in separate sessions and with differently worded prompts. Labeller A applied the codebook. Labeller B audited the checker by searching the code itself (Figure~\ref{fig:prompts}). Neither saw the other's labels. Where they disagreed, a third LLM pass read both labels and the evidence and proposed one label. All of these outputs were proposals. They made human review faster, because a reviewer could check the cited evidence instead of searching a whole codebase, but none was used as a final label.

\paragraph{Role 3: Writing analysis scripts.}
We used Claude as a coding assistant to help write the data-processing scripts, the path extractor, the import checker, and the scripts that compute statistics and draw figures. The first author specified what each script should do, read the code, and checked its output by hand on samples and small worked examples. The review samples also serve as an end-to-end check, because they compare script output with the repositories themselves. The research design, the choice of measures and tests, the analysis and the interpretation of results were done by the authors.

\paragraph{What we did not use LLMs for.} No LLM produced a number in our results directly. The path extractor, the claim-history replay, the import checker and the statistical tests are ordinary deterministic programs: the same input always gives the same output. 

\paragraph{Reporting.}
We follow the community guidelines for empirical studies involving LLMs \citep{baltes2027llmguidelines}. Figure~\ref{fig:prompts} shows the prompt excerpts. The replication package contains the prompt templates for every LLM step, the codebook the prompts applied, the inputs and raw outputs of every LLM step, the reasons each labeller gave, and the agreement figures.

\begin{figure}[t]
\centering
\small
\fbox{\parbox{0.92\textwidth}{
\textbf{Labeller A:} \textit{CODEBOOK (apply exactly): Only static import/include/require/use statements in production code count. Do NOT count: dependencies declared only in build or manifest files; dynamic or runtime loading; [\ldots] HOLDS: no counted import breaks the rule as its author meant it. MISSED\_VIOLATION: at least one counted import breaks it (cite file:line).}\\[4pt]
\textbf{Labeller B:} \textit{Audit the checker by searching the code yourself. [\ldots] What never counts: entries in build/manifest files; runtime or dynamic loading [\ldots] HOLDS = nothing that counts breaks the rule. MISSED\_VIOLATION = something that counts breaks it (give file:line).}
}}
\caption{Excerpts of the two LLM labelling prompts for rules that held.}
\label{fig:prompts}
\end{figure}

\subsection{Robustness checks}
\label{sec:robust}

We ran three simple checks to see whether our main results depend on choices we made.
\begin{itemize}
\item \emph{Projects with many claims (RQ1).} Some projects contribute many path claims. We therefore recomputed the interval for the missing rate by resampling whole projects (5,000 bootstrap samples) \citep{efron1993bootstrap}, and combined it with the uncertainty in the checker's precision.
\item \emph{Review and detection choices (RQ3).} We repeated the commit comparison (i) with all flagged breaks, before human review, and (ii) with agent-assisted commits defined by commit-message traces only.
\item \emph{Single projects (RQ3).} Few projects had any break, so we repeated the comparison leaving out one project with a confirmed break at a time.
\end{itemize}

\subsection{Human review}
\label{sec:checks}

Every automatic step can make mistakes. We therefore checked a sample of the output of each step by hand. Table~\ref{tab:checks} summarizes all human review in the study. Each review followed the same four stages.

\paragraph{Stage 1: Sample.}
We drew a sample from the step's output. Where a group was small, we took all of it. For example, we reviewed all 51 rules that were broken when written.

\paragraph{Stage 2: LLM proposals.}
The two LLM labellers proposed a label and a reason for every sampled item (Section~\ref{sec:llmuse}, Role~2).

\paragraph{Stage 3: The first author decides.}
The first author reviewed every sampled item. For each one, they read the proposed label and reason, checked the evidence in the repository, and set the final label. The first author also reviewed all 184 structured rules. Where the first author's judgment differed from the LLM proposals, the first author's label stood. All results in this paper use the first author's final labels.

\paragraph{Stage 4: A second reviewer verifies the key checks.}
An external reviewer, not otherwise involved in the study, verified the checks that underlie our main findings and that require judging what a rule's author meant:
\begin{itemize}
\item all 51 rules broken when written;
\item all 40 sampled rules that held;
\item all 35 commit-rule pairs flagged as adding a break;
\item 40 structured rules, sampled at random from the held rules not already covered by the two rule checks above.
\end{itemize}
For each item, the second reviewer saw the first author's label and the LLM's reason, opened the evidence, and marked the label as correct or not. Where they disagreed, they gave the label they thought was correct. We report the share of labels the second reviewer accepted, and Cohen's $\kappa$ between the two reviewers, in Table~\ref{tab:checks}. We then discussed each disagreement, and the first author set the final label. No LLM took part in this step.

The second reviewer accepted 49 of 51 labels for rules broken when written, 39 of 40 for rules that held, and 34 of 35 for flagged commits ($\kappa$ between 0.89 and 0.94). We discussed the four disagreements. In one, the review sheet lacked evidence that the data hold: the nine rule-breaking imports behind the rule. In one, both reviewers agreed that a structured rule was reversed, and the codebook's decision rule for that case kept the original label. In the other two, the case turned on what a rule's author meant. In all four, the first author kept the original label, so no final label changed, and no result had to be recomputed.

For the structured rules, the second reviewer accepted 34 of 40. Five of the six disagreements came from two fields that the review sheet did not show. Four rules apply to each sibling folder separately, so a module that imports its own files does not break them. One rule exempts a target folder. The sixth rule's structured form captures only part of its meaning. It is one of the rules that held, so its check may miss a break, in line with the lower bound we report in RQ3.

The other checks concern facts that can be observed directly, such as whether a folder exists or whether a commit message has an agent co-author line. For these, the two LLM labellers agreed on 94\% to 100\% of items, and the first author's review settled the rest. We did not give them to the second reviewer.

\paragraph{The codebook and its pilot.}
All reviewers used one codebook, which is in the replication package. Its main rule is to judge what the \emph{author of the context file meant}, using the surrounding text and the repository at the linked commit. For rules that held, a pilot round with the LLM labellers showed low agreement (78\%, $\kappa = 0.39$). The disagreements came from cases the codebook did not yet cover, such as dependencies declared only in build files, and dynamic loading. We added decision rules for these cases. Both labellers then relabelled all 40 rules, and their agreement rose to 95\% ($\kappa = 0.72$).

\begin{table}[t]
\centering
\caption{All human review in the study. \emph{LLM A vs.\ B}: agreement between the two LLM labellers that proposed labels. \emph{Second reviewer}: items verified, share of the first author's labels accepted, and Cohen's $\kappa$ between the two reviewers. Final labels and results are the first author's.}
\label{tab:checks}
\small
\setlength{\tabcolsep}{3.5pt}
\begin{tabularx}{\textwidth}{@{}>{\raggedright\arraybackslash}Xrrrrrr>{\raggedright\arraybackslash}p{3.1cm}@{}}
\toprule
 & & \multicolumn{2}{c}{LLM A vs.\ B} & \multicolumn{3}{c}{Second reviewer} & \\
\cmidrule(lr){3-4}\cmidrule(lr){5-7}
What we checked & Items & Agree & $\kappa$ & Items & Accepted & $\kappa$ & Result (first author)\\
\midrule
\multicolumn{8}{@{}l}{\emph{Rules (Step 4)}}\\
Structured rules & 184 & \multicolumn{2}{c}{one proposal} & 40 & 85\%\textsuperscript{d} & n/a\textsuperscript{c} & All 184 reviewed\\
Rules broken when written & 51 & 84\% & 0.66 & 51 & 96\% & 0.92 & 33 confirmed\\
Rules that held\textsuperscript{a} & 40 & 95\% & 0.72 & 40 & 98\% & 0.89 & 5 had a missed break\\
\midrule
\multicolumn{8}{@{}l}{\emph{Commits (Step 5)}}\\
Commits flagged as adding a break & 35\textsuperscript{b} & 83\% & 0.66 & 35 & 97\% & 0.94 & 14 commits confirmed\\
Exposed commits not flagged & 100 & 100\% & n/a\textsuperscript{c} & & & & No missed break\\
Agent-assisted labels & 150 & 95\% & 0.88 & & & & 96 of 97 agent labels confirmed; 3 of 53 others had a trace\\
\midrule
\multicolumn{8}{@{}l}{\emph{Path claims (Step 2)}}\\
Missing path claims & 300 & 94\% & 0.87 & & & & 42\% (context files) and 29\% (READMEs) are real errors\\
Found path claims & 100 & 97\% & n/a\textsuperscript{c} & & & & 99 correct\\
\bottomrule
\end{tabularx}
\par\smallskip
\begin{minipage}{\textwidth}\footnotesize
\textsuperscript{a}LLM figures after the pilot round (pilot: 78\%, $\kappa=0.39$). \textsuperscript{b}Commit-rule pairs (30 commits). \textsuperscript{c}$\kappa$ is not informative when nearly all items share one label. \textsuperscript{d}Five of the six disagreements came from fields the review sheet did not show (see text).
\end{minipage}
\end{table}

\section{Results}
\label{sec:results}

\subsection{RQ1: How accurate are the claims?}
\label{sec:rq1}

\paragraph{Context files make many more path claims than READMEs.}
At the snapshot, context files contain 41,090 path claims in 3,227 projects, a median of 9 per project. Only 521 READMEs in the same projects make any path claims (1,857 claims, a median of 1). Developers describe their project's layout for agents far more than for human readers.

\begin{figure}[t]
\centering
\includegraphics[width=0.75\textwidth]{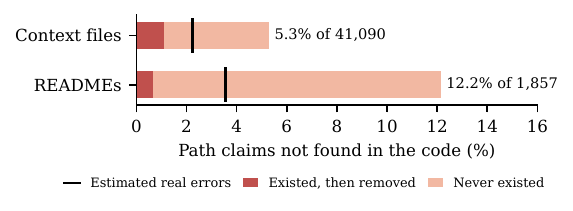}
\caption{Path claims missing at the snapshot. The black line shows the estimated real errors.}
\label{fig:rq1}
\end{figure}

\paragraph{About 5\% of context-file claims are missing, but fewer than half of these are real errors.}
Our checker does not find 2,182 context-file claims (5.3\%) and 226 README claims (12.2\%) (Figure~\ref{fig:rq1}). These raw rates include extraction noise. Table~\ref{tab:validation} shows the labels for the sampled missing claims.
\begin{itemize}
\item In context files, 42\% of missing claims are real errors (95\% interval 35\% to 50\%).
\item In READMEs, only 29\% are (22\% to 38\%). READMEs often show paths in the user's own project, which are not errors.
\end{itemize}
This gives an estimated real error rate of 2.2\% for context files (1.9\% to 2.6\%) and 3.5\% for READMEs (2.7\% to 4.6\%). In other words, about 98\% of context-file path claims point to real code. Of the 100 found claims we checked, 99 were correct, so lenient matching rarely accepts a wrong claim.

The low precision is expected. Our extractor is generous on purpose, so that it misses few claims. The price is that it also picks up paths that are not claims about the repository. This is why the adjusted rate, not the raw rate, is the number to trust.

In the 453 projects where both kinds of documents make path claims, the raw missing rate is much lower for context files (4.9\% vs.\ 11.2\%). After adjusting for precision, however, the two are close (about 2.1\% vs.\ 3.3\%). We therefore see the two as similar, rather than one as more accurate.

\paragraph{The result does not depend on a few projects.}
Most projects (72\%) have no missing claim at all, and the ten projects with the most missing claims account for 11\% of them. Resampling whole projects (Section~\ref{sec:robust}) gives a 95\% interval of 4.8\% to 5.8\% for the raw missing rate. Combined with the uncertainty in precision, the estimated real error rate stays between 1.8\% and 2.7\%.

\begin{table}[t]
\centering
\caption{Labels for the sampled missing path claims.}
\label{tab:validation}
\small
\begin{tabular}{@{}lrr@{}}
\toprule
Label & Context files & READMEs\\
\midrule
\textsc{Wrong path} (real error) & 76 & 35\\
\textsc{Not a repository path} & 53 & 65\\
\textsc{Placeholder or example} & 20 & 9\\
\textsc{Parse error} & 18 & 2\\
\textsc{Generated or ignored} & 13 & 9\\
\midrule
Precision (share of real errors) & 0.42 & 0.29\\
\bottomrule
\end{tabular}
\end{table}

\paragraph{Raw rates are similar across files and languages.}
The raw missing rate ranges from 4.5\% (Python) to 7.8\% (PHP) across languages, and from 5.0\% to 6.1\% across the main file types (Table~\ref{tab:breakdown}). Files first committed with an agent trace show a similar rate to other files (4.9\% vs.\ 5.5\%). We did not estimate precision for each subgroup, so small differences may be noise.

\paragraph{Real errors split between paths that never existed and paths that went stale.}
Of the 2,182 missing paths, 1,737 (80\%) never existed while the claim was in the file. The other 445 (20\%) existed when the claim was written and were removed later. The two groups differ in precision. Nearly all sampled stale claims are real errors (30 of 32), but only 31\% of never-valid claims are. Weighting by precision, about 56\% of real errors are claims that were wrong from the start, and 44\% are claims that went stale.

Claims that were wrong from the start include:
\begin{itemize}
\item files renamed before the claim was written (\texttt{skill.md} vs.\ \texttt{SKILL.md});
\item folders of a planned layout that was never built;
\item files that seem invented.
\end{itemize}
In 27 of the 76 real errors we labelled, a file with the same name exists elsewhere in the repository. This suggests a move or rename.

\paragraph{Many dependency rules do not hold.}
At the snapshot, our checker finds rule-breaking imports for 54 of the 184 rules (29\%), with 906 such imports in total. Human review confirmed 32 of these rules. So at least one rule in six does not hold. We return to these rules in RQ3, because nearly all of them were already broken on the day they were written.

\finding{\textbf{Summary of RQ1 findings.} Context files make far more path claims than READMEs. About 5\% of these claims point to paths we cannot find. After human review, we estimate that only about 2.2\% are real errors, similar to READMEs (3.5\%). Slightly more of the real errors were wrong from the start than went stale. In contrast, at least one in six dependency rules is broken at the snapshot.}

\subsection{RQ2: Do claims stay accurate as the code changes?}
\label{sec:rq2}

\paragraph{Few claims go stale in the observed window.}
We follow 44,732 context-file claims that were valid when written (Table~\ref{tab:rq2}). Each was followed for a median of 1.7 months. In that time, 1,151 of them (2.6\%) went stale because their path was removed. This is 0.99 stale claims per 100 claim-months. For README claims, the rate is 1.74 per 100 claim-months (75 of 1,953 claims). Within the 441 projects that have both kinds of claims, the rates are nearly equal (1.86 vs.\ 1.83). In 12\% of projects, at least one context-file claim went stale.

\begin{table}[t]
\centering
\caption{What happened to path claims that were valid when written.}
\label{tab:rq2}
\small
\begin{tabular}{@{}lrr@{}}
\toprule
 & Context files & READMEs\\
\midrule
Claims valid when written & 44,732 & 1,953\\
\quad still valid at the snapshot & 38,805 & 1,607\\
\quad deleted from the file while still valid & 4,776 & 271\\
\quad went stale & 1,151 (2.6\%) & 75 (3.8\%)\\
\qquad fixed & 706 (61\%) & 63 (84\%)\\
\qquad unfixed at the snapshot & 445 (39\%) & 12 (16\%)\\
Stale claims per 100 claim-months & 0.99 & 1.74\\
Median days from writing to staleness & 20 & 9\\
Median days from staleness to fix & 0 & 0\\
Median age of unfixed stale claims (days) & 59 & 82\\
\bottomrule
\end{tabular}
\end{table}

\begin{figure}[t]
\centering
\includegraphics[width=0.75\textwidth]{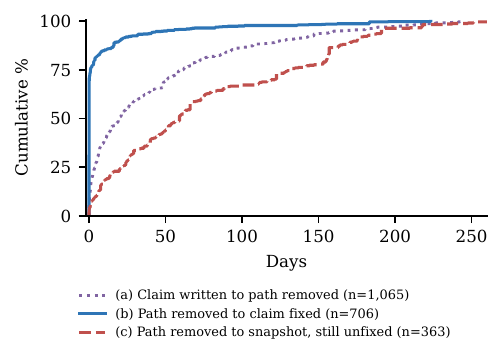}
\caption{Timing of staleness in context files (cumulative): (a) time from writing to staleness, (b) time from staleness to fix, (c) age of unfixed stale claims.}
\label{fig:rq2}
\end{figure}

\paragraph{Claims go stale soon after they are written.}
Figure~\ref{fig:rq2} (curve a) shows how long claims stayed valid before their path was removed. We know the removal commit for 1,065 of the stale context-file claims. For these, the median time from writing the claim to removing the path was 20 days. Half fell between 4 and 61 days. About 34\% went stale within a week and 59\% within a month. Only 16\% lasted more than three months. README claims show a similar pattern (median 9 days). Staleness is therefore not a slow decay. It happens early, perhaps because the code is still being reorganized when the file is written.

\paragraph{Stale claims are often fixed at once.}
Of the 1,151 stale context-file claims, 706 (61\%) were removed or corrected by the snapshot. Most fixes were immediate (Figure~\ref{fig:rq2}, curve b). In 61\% of fixes, the claim changed at the same time as the code, which in practice means in the same commit. In 75\% of fixes, the claim was fixed on the same day. Only 8\% took more than 30 days. Developers, or their agents, seem to update the context file as part of the change that moves the code. Another 4,776 valid claims were deleted for other reasons, so these files are actively edited.

\paragraph{Stale claims that are not fixed quickly stay stale.}
The other 445 stale claims (39\%) were still in place at the snapshot, a median of 59 days after their path was removed (Figure~\ref{fig:rq2}, curve c). At the snapshot, 6.8\% of projects had at least one such claim. In short, a stale claim is usually fixed with the change or not at all. This fits with the 38\% of context files that were never edited after they were created.

\paragraph{Some kinds of claims are kept up better than others.}
Table~\ref{tab:breakdown} breaks the results down.
\begin{itemize}
\item Claims in structure or architecture sections go stale about half as fast as claims elsewhere (0.72 vs.\ 1.53 per 100 claim-months).
\item Folder trees go stale more slowly than inline paths, but their stale entries are fixed less often (55\% vs.\ 63\%).
\item Stale claims in Copilot instruction files are fixed far less often (43\%) than those in \texttt{AGENTS.md} (73\%) or \texttt{CLAUDE.md} (59\%).
\end{itemize}

\begin{table}[t]
\centering
\caption{Context-file path claims by file type, section, form and language.}
\label{tab:breakdown}
\small
\begin{tabular}{@{}lrrrrr@{}}
\toprule
 & Claims & Projects & Missing & Stale\textsuperscript{a} & Fixed\textsuperscript{b}\\
\midrule
\texttt{CLAUDE.md} & 15,219 & 1,206 & 5.0\% & 0.95 & 59\%\\
\texttt{AGENTS.md} & 16,982 & 1,435 & 5.1\% & 1.20 & 73\%\\
Copilot instructions & 7,838 & 768 & 6.1\% & 0.77 & 43\%\\
Other files\textsuperscript{c} & 1,051 & 130 & 7.5\% & 0.98 & 58\%\\
\midrule
Structure/architecture section & 26,004 & 2,467 & 4.6\% & 0.72 & 59\%\\
Other sections & 15,086 & 2,706 & 6.5\% & 1.53 & 64\%\\
\midrule
Inline path & 30,608 & 3,142 & 5.0\% & 1.08 & 63\%\\
Folder tree & 10,482 & 671 & 6.2\% & 0.70 & 55\%\\
\midrule
TS/JS/Python & 21,389 & 1,625 & 5.0\% & 0.82 & 56\%\\
Other seven languages & 19,701 & 1,602 & 5.7\% & 1.17 & 65\%\\
\bottomrule
\end{tabular}
\par\smallskip
{\footnotesize Missing: raw rate. \textsuperscript{a}Per 100 claim-months. \textsuperscript{b}Share of stale claims fixed. \textsuperscript{c}\texttt{.claude/CLAUDE.md}, \texttt{GEMINI.md}, \texttt{.cursorrules}.\par}
\end{table}

\finding{\textbf{Summary of RQ2 findings.} Few context-file claims go stale in the observed window (0.99 per 100 claim-months, similar to READMEs in the same projects). Those that do mostly go stale within a month of being written. More than half of stale claims are fixed, most in the same commit that changes the code. Claims that are not fixed at once tend to stay stale for months.}

\subsection{RQ3: What happens to rules after they are written?}
\label{sec:rq3}

\paragraph{Rules are mostly about module boundaries.}
The 184 checkable rules come from 115 projects in all ten languages, most often TypeScript, Python, Go and Java. Of these rules:
\begin{itemize}
\item 126 restrict imports between parts of the project, for example ``the \texttt{core} layer cannot import \texttt{api}, \texttt{cli} or \texttt{utils}'';
\item 58 restrict which parts may use an external library, for example ``never import \texttt{tokio} directly outside of \texttt{runtime/}''.
\end{itemize}
Most rules (176) forbid an import. The other 8 allow an import only from certain places. Table~\ref{tab:rules} gives examples.

\begin{table}[t]
\centering
\caption{Example rules and their rule-breaking imports.}
\label{tab:rules}
\small
\begin{tabularx}{\textwidth}{@{}>{\raggedright\arraybackslash}Xlrr@{}}
\toprule
Rule (shortened) & Lang. & At writing & At snapshot\\
\midrule
Do not import MUI; we are actively removing it & TS & 132 & 74\\
Do not import \texttt{encoding/json} in business code & Go & 107 & 107\\
\texttt{hopsworks\_common} must not import \texttt{hsfs}, \texttt{hsml} & Py & 127 & 124\\
\texttt{pwndbg/dbg\_mod/} never imports \texttt{aglib} & Py & 30 & 31\\
Never fall back to \texttt{serde\_json}; use Norito & Rust & 24 & 24\\
No dependencies on legacy \texttt{pkg/og/} (being phased out) & Go & 21 & 21\\
Protocol modules do not depend on each other & Java & 0 & 0\\
\texttt{io} only depends on \texttt{core} & C++ & 0 & 0\\
\bottomrule
\end{tabularx}
\par\smallskip
{\footnotesize Numbers are rule-breaking imports when the rule was written and at the snapshot.\par}
\end{table}

\begin{figure}[t]
\centering
\includegraphics[width=0.75\textwidth]{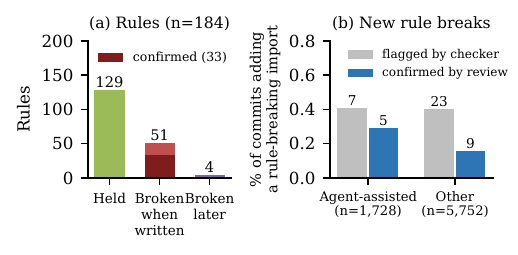}
\caption{(a) Status of the 184 dependency rules. (b) Exposed commits that add a rule break, by commit group.}
\label{fig:rq3}
\end{figure}

\paragraph{Broken rules were broken from the start.}
Figure~\ref{fig:rq3}a shows the status of each rule.
\begin{itemize}
\item 129 rules held from the day they were written to the snapshot.
\item 51 rules were already broken on the day they were written. Of these, 50 were still broken at the snapshot.
\item Only 4 rules held when written and were broken later.
\end{itemize}
Human review confirmed 33 of the 51 rules that were broken when written. In the other 18, the check mostly did not match the rule's meaning. Common reasons were exceptions stated in a lint configuration, or type-only imports. Counting only confirmed cases, 18\% of rules were broken on the day they were written, in 21\% of the projects with rules. This is likely a lower bound. In a sample of 40 rules that our checker found intact, 5 had a break that it missed.

Some broken rules say so openly. One file notes that \texttt{pkg/og/} ``is legacy code being phased out''. Another says that ``we are actively removing'' Material UI (MUI). Other files give no hint that the rule is not yet true. In these examples, the rules describe a target architecture rather than the current one. Rules about external libraries were broken from the start more often than rules between internal modules (14 of 58, or 24\%, vs.\ 19 of 126, or 15\%, confirmed). Several of the library rules describe a move away from a library.

\paragraph{The rule-breaking code is mostly older than the rule.}
In 26 of the 33 confirmed cases, every rule-breaking import was already in the code before the project's first context file existed. In most cases, then, nobody broke the rule after it was written. The rule was written on top of code that did not follow it. Most of these rules (24 of 33) were written in commits without an agent trace. In three projects, agent-assisted commits wrote both a rule and some of the code that broke it.

\paragraph{New rule breaks are rare.}
After the rules were written, we found 7,480 exposed commits in 96 projects. Of these, 1,728 (23\%) were agent-assisted, in 56 projects. Our checker flagged 30 commits as adding a rule break: 7 agent-assisted and 23 other, or 0.4\% of each group (Figure~\ref{fig:rq3}b). Human review confirmed 14 of them: 5 agent-assisted commits and 9 other commits. The confirmed agent-assisted breaks did one of three things:
\begin{itemize}
\item imported a GUI toolkit into a presenter layer;
\item imported a UI library directly instead of through the project's wrapper;
\item bypassed required import aliases.
\end{itemize}
We also checked 100 exposed commits that were not flagged, and found no missed break.

\paragraph{Agent-assisted commits broke rules more often, but the evidence is early.}
We asked an exploratory question here: do commits that agents helped make break the rules more often? The answer below is an early signal, not a test of a firm hypothesis. Counted per 1,000 exposed commits, agent-assisted commits added 2.9 confirmed breaks and other commits 1.6, about 1.8 times as many. Repairs point the same way. Of the 29 commits that removed rule-breaking imports, only 2 were agent-assisted, against about 7 expected at equal rates. Comparing the groups within projects, the odds ratio is 2.6, with a wide 95\% interval from 0.7 to 10.2 ($p = 0.21$). Only 8 of the 51 projects with both groups had any confirmed break. The data are therefore compatible both with no difference and with a several times higher rate in agent-assisted commits. The lack of a significant result does not show that the two groups behave the same.

\paragraph{The direction holds under other choices.}
We repeated the comparison in three ways (Section~\ref{sec:robust}).
\begin{itemize}
\item With all 30 flagged breaks, before human review, the odds ratio is 2.4 (95\% interval 0.9 to 6.5, $p = 0.13$).
\item With agent-assisted commits defined by commit-message traces only, it is 2.7 (0.7 to 10.3, $p = 0.20$).
\item Leaving out one project at a time gives odds ratios between 1.9 and 5.4. In one case, $p$ falls just below 0.05. In all others, it stays above 0.17.
\end{itemize}
The direction is stable: agent-assisted commits never break rules less often. How large any difference is remains uncertain, because one project can move the estimate a lot.

\paragraph{How much data a firm answer would need.}
A rough power estimate, which ignores clustering by project, shows the scale of the gap. Detecting a difference of 1.8 times with 80\% power would need about 54,000 exposed commits, roughly seven times our sample, or about 100 confirmed breaks. We therefore report the comparison as an early signal for larger studies to test, not as a finding.

\finding{\textbf{Summary of RQ3 Findings.} Most broken rules were broken on the day they were written, usually by code that predates the rule, and they stayed broken. New breaks were rare. In an exploratory comparison, agent-assisted commits broke rules about 1.8 times as often as other commits (0.29\% vs.\ 0.16\%) and repaired them less often. The direction held in every robustness check, but the difference rests on 14 breaks and is not statistically significant. It is an early signal for larger studies to test, not evidence of a difference or of its absence.}

\section{Discussion}
\label{sec:discussion}

\subsection{Good maps, aspirational rulebooks}

As \emph{maps}, context files are mostly accurate. An estimated 98\% of path claims point to real code. This is similar to READMEs in the same projects, and similar across languages. When code moves, the map is often updated in the same commit. This is more upkeep than the documentation literature would lead us to expect \citep{aghajani2019issues,wen2019comments,migliorini2024views}. One reason may be that an agent fails visibly when the map is wrong. Another may be that agents edit the file themselves while making the change.

As \emph{rulebooks}, context files are less reliable. Almost one in five rules was broken on the day it was written, and these rules stayed broken. In classic terms, the described and the implemented architecture disagreed from the start \citep{perry1992foundations}. This is not the familiar story of erosion, where a sound design decays over time \citep{desilva2012erosion}. In the examples we inspected, it looks more like a team writing down the architecture it wants. One reason may be that rules are rarely checked: developers report that conformance checks are seldom done, let alone automated \citep{melo2016perceptions,caracciolo2016monitoring}, so nothing flags a rule that is broken on the day it is written.

This matters for agents in a specific way. A human newcomer who reads ``do not import \texttt{encoding/json} in business code'' and then finds it imported in 107 places will ask a colleague. An agent sees two conflicting signals: the rule in its instructions, and many counter-examples in the code it reads. In our data, new rule breaks were rare in both commit groups. Still, the gap between the stated and the actual architecture is the kind of hidden inconsistency that Santos et al.~\citep{santos2026smells} describe as a configuration smell.

\subsection{Wrong from the start vs.\ stale later}

The history analysis shows where errors come from. Dependency rules are often wrong from the day they are written. For path claims, the split is closer. Slightly more of the real errors were wrong from the start than went stale, and stale claims are often fixed together with the code change. This adds to the idea of context rot \citep{treude2026contextrot}: claims do not only rot over time, they can also be wrong or aspirational from the start. Tools that check a context file at the moment it is written may therefore catch more problems than tools that watch for drift later. The broken rules also do not look like reactions to agent mistakes. Practitioners say they mostly add rules to correct AI errors \citep{cai2026rules}, but in our data the rule-breaking code was usually older than the context file.

The rules that were broken from the start often describe a migration, such as moving off a legacy package, a UI library or a serialization library. We read these, cautiously, as statements of a target architecture. They tell the agent what \emph{new} code should look like, while old code does not yet comply. Most files do not say so, however. The few that do, for example by stating that a dependency is ``being phased out'', suggest a simple fix: state the target, the current exceptions, and the direction of change. Architecture research separates the \emph{prescriptive} architecture (what should be) from the \emph{descriptive} architecture (what is) \citep{perry1992foundations}. Context files mix both in one list. Such rules can also be read as architectural technical debt \citep{verdecchia2021atd}: like debt that developers admit in code comments \citep{sierra2019satd}, they record a gap between the intended and the actual architecture, but without marking it as one.

\subsection{Agents and rules}

The exploratory part of RQ3 points toward more breaks, and fewer repairs, in agent-assisted commits: about 1.8 times as many breaks per commit. But it rests on 14 confirmed breaks, and only 8 projects with both commit groups had any. Two readings fit the data. Agent-assisted commits may mostly respect explicit negative rules, as controlled studies suggest \citep{gloaguen2026agentsmd,zhang2026guardrails}. Or there is a real difference that our data are too small to show; a firm answer would need roughly seven times as many exposed commits. Either way, the larger gap in our data lies between the rules and the code, since the broken rules were usually broken before they were written.

\subsection{Human-AI collaboration in software architecture}

In human-agent teamwork, common ground lets each side predict and direct the other \citep{klein2004teamplayer}. A context file is a very explicit part of this common ground. It is written down, versioned, and handed to the agent every session. It can be seen as a \emph{boundary object} \citep{star1989boundary}: a shared artifact that both people and agents use, each in their own way.

Read this way, the map mostly matches the system (RQ1) and often follows the code quickly (RQ2). The weak spot is the content of the rules. The written rules often state what the team \emph{intends}, while the code shows what it \emph{does}. A human teammate often knows that a migration is under way. An agent does not, unless someone writes it down.

Storey's triple debt model offers one lens on these results \citep{storey2026tripledebt}. It separates three kinds of debt: technical debt in the code, cognitive debt in the team's shared understanding, and intent debt in missing or unclear goals and constraints. Path claims suggest that structural knowledge can be written down and kept current at low cost. Aspirational rules, however, are a concrete form of intent debt. The intent is captured, but it is not marked as a goal. An accurate file also says little about what the team itself understands. Cognitive debt may therefore shift toward intent debt rather than disappear.

This points to a wider question. Research on teams separates knowledge that members hold individually from knowledge they \emph{share}. Only shared knowledge supports coordination \citep{cannonbowers1993smm,mohammed2010metaphor}. A context file is often written by one person, sometimes with an agent's help. It may reflect the whole team's understanding, its author's view, or a mix of actual and intended architecture. Naur~\citep{naur1985theory} argued that the ``theory'' of a program cannot be fully written down. How the knowledge written for agents relates to the understanding held by the team is still an open question. In our view, it is an important one for human-AI teams that aim to stay aligned on architecture.

\subsection{Using LLMs as research tools in mining studies}
\label{sec:llmdisc}

LLMs are now common tools in mining software repositories. A recent survey found 85 such studies between 2017 and 2025 \citep{romeroarjona2026rise}. Classification was the most common use. In many studies, the language model was not the subject of the research but a helper that turned raw repository data into structured input for later analysis \citep{romeroarjona2026rise}. Our study uses LLMs in exactly this helper role. A review of 50 empirical software engineering studies that used LLMs found the same pattern. The main benefits reported were efficiency and scale. The main limitations were hallucination, inconsistency, sensitivity to prompts and poor reproducibility, and reporting was often incomplete \citep{gomes2026llmese}. The field now has guidelines for reporting such use \citep{baltes2027llmguidelines}, frameworks that link each step of an LLM-based mining study to its threats and mitigations \citep{demartino2025primes}, and proposals to treat LLM labelling as a measurement process, with its reliability and drift reported like any other instrument \citep{imran2026olaf}. Here we share what the LLMs gave us, what we learned, which threats they bring, and how we handled them.

\paragraph{What LLMs made possible.}
Two parts of our study would have been hard without LLMs. First, the keyword filter returned 1,645 candidate sentences from 875 projects. Each one had to be read, judged, and, if it was a rule, mapped to real folders in its repository. An LLM did the first pass, and a person reviewed every rule it produced. Second, reviewing a rule means searching a large codebase for imports that break it. The LLM labellers did this search and returned a label with a file and line as evidence. Our reviewers could then check that evidence directly instead of searching from scratch. This matches findings elsewhere. In text annotation, an LLM matched or beat crowd workers at a fraction of the cost \citep{gilardi2023chatgpt}. For software engineering artifacts, LLM labels reached agreement close to human agreement on some tasks, but not all \citep{ahmed2025annotators}.

\paragraph{Disagreement between LLMs showed us where the hard cases were.}
Ahmed et al.~\citep{ahmed2025annotators} suggest that agreement between two models can predict whether a labelling task suits LLMs at all. Our data fit this idea. The two LLM labellers agreed on 94\% of missing paths and 95\% of agent labels. They agreed less often when labelling the rules that were already broken on the day they were written (84\%, $\kappa = 0.66$). There, the label depends on what the rule's author meant, not only on what the code contains. These are also the checks we gave to a second human reviewer. The LLM pilot on rules that held was even more useful. Its low agreement ($\kappa = 0.39$) showed that our codebook did not yet cover build-file dependencies or dynamic loading. In effect, the two labellers tested our codebook before the codebook reviewers used it. Others report similar experiences in qualitative coding of text. Dunivin et al.~\citep{dunivin2026selfreflection} had a second LLM re-check the labels of a first one on developer emails, and adapted their codebook to the errors a human audit found. In thematic analysis of software engineering interviews, experts often preferred LLM codes but found that the models missed deeper interpretations \citep{martinezmontes2025thematic}. Our experience adds a setting where the labels are checks against code, so each proposal comes with evidence, such as a file and line, that a reviewer can open and verify.

\paragraph{Threats LLMs bring.}
Using LLMs as research tools brings its own threats to validity \citep{sallou2024breaking}. Table~\ref{tab:llmthreats} lists the ones that apply to our study.
\begin{itemize}
\item \emph{Shared errors.} Two LLM labellers can make the same mistake. Across more than 350 models, Kim et al.~\citep{kim2025correlated} found that model errors are strongly correlated, even across providers and architectures. In a recent census of agent instruction files, LLM annotators agreed more with each other than with the human reviewer \citep{zarzycki2026census}. Our two labellers used the same model with different prompts, so shared errors are even more likely. Agreement between them therefore shows consistency, not correctness.
\item \emph{Non-determinism.} The same prompt can give different output on different runs, and a temperature of zero does not remove this \citep{ouyang2025nondeterminism}. Rerunning our LLM steps could change some proposed labels or structured rules.
\item \emph{Closed and changing models.} Commercial models change over time and may be retired \citep{sallou2024breaking}. The public repositories we study may also be part of their training data.
\item \emph{Anchoring of human reviewers.} A reviewer who reads an LLM's label and reason may lean toward it. AI-generated analyses can look correct and still lead to wrong conclusions, and people find such errors hard to spot \citep{gu2024analysts}.
\item \emph{Errors in LLM-written scripts.} The same risk applies to analysis code written with LLM help. A script can run and produce plausible numbers that are wrong \citep{gu2024analysts}.
\item \emph{LLMs judging text written for LLMs.} Our study has an unusual loop. We used LLMs to study files that are written for LLMs, and sometimes by them. A model may read such text the way the agent does, and share its blind spots.
\end{itemize}

\begin{table}[t]
\centering
\caption{Threats from using LLMs in this study, and our mitigation strategy for each.}
\label{tab:llmthreats}
\small
\begin{tabularx}{\textwidth}{@{}>{\raggedright\arraybackslash}p{2.7cm}>{\raggedright\arraybackslash}X@{}}
\toprule
Threat & Mitigation Strategy\\
\midrule
Shared errors between LLM labellers & LLM labels were proposals only; the first author set every final label; a second reviewer verified the key checks\\
Non-determinism & Final labels and rules are fixed human decisions; all raw LLM outputs are archived\\
Closed and changing models & Model named; codebook and raw outputs shared; no result depends on rerunning a model\\
Anchoring of reviewers & Reviewers checked the evidence in the repository, not only the LLM's reason; written codebook; agreement with a second reviewer reported\\
Errors in LLM-written scripts & First author specified, read and tested each script; review samples check output against the repositories\\
LLMs judging text written for LLMs & Deterministic import checker decides whether a rule holds; human judgment of what each rule means\\
\bottomrule
\end{tabularx}
\end{table}

\paragraph{How we kept people in charge.}
Our main safeguard was a division of labour. LLMs proposed, and people decided. Three choices made this work. First, we kept the measurement itself deterministic. The path checker, history replay and import checker are ordinary programs, so the core numbers do not depend on a model. Second, every LLM output that fed a result passed through human review. The first author set every final label, and a second reviewer verified the checks behind the main findings. Third, we report each use. We name the model, describe its roles, and share the codebook, raw outputs and agreement figures, as current guidelines ask \citep{baltes2027llmguidelines}. Ernst and Treude~\citep{ernst2026silverbullet} warn that GenAI is no silver bullet for qualitative work in software engineering, and that its use must fit both the data and the research strategy. In our study, it fit best where the task was well defined and the evidence could be checked, such as finding an import in a file. It fit least where the task required judging intent, which is where we relied most on people. Experienced researchers reach a similar view: LLMs can support interpretive analysis, but not replace it, and continuous human oversight is needed \citep{ornelas2026thematic}.

\subsection{Implications}
\label{sec:implications}

\textit{For developers.}
Treat the architecture section of a context file like code. Check new path claims when you write them. Mark rules that are goals rather than facts, for example a dependency that is being phased out. Then both the agent and your teammates know the difference. Where a rule matters, enforce it with a linter, as some projects already do (``enforced by ESLint \texttt{no-restricted-imports}''). Such checks work in practice: when a conformance check ran in continuous integration, the violations it found were fixed before each release \citep{pinto2017archci}. \textit{For tool builders.} A simple checker that resolves path claims and runs stated dependency rules at commit time could catch many of the problems we found. It should also run whenever a context file changes. Tools like ContextCov \citep{sharma2026contextcov} fit this need. Since many rules are broken from day one, however, such tools must support known exceptions and migration states. \textit{For researchers.}
Studies that measure whether context files help agents should also check whether the content is true. Wrong or aspirational content may hide real effects. Mining studies of agent activity should look beyond commit co-author lines. In our data, pull-request traces found about a quarter more agent-assisted commits. Studies that use LLMs as research tools should describe each use and how humans checked it, as we discuss in Section~\ref{sec:llmdisc}. Finally, the architectural claims written for agents are a large, fresh source of data on intended architecture, stated right next to the code they govern \citep{soliman2025msr4sa}.

\section{Threats to Validity}
\label{sec:threats}

\paragraph{Construct validity.}
We define architecture through path claims and dependency rules. Other architectural content, such as data flow, is out of scope. Our extractor produces false ``missing'' results. We measured its precision on a sample and report adjusted rates with intervals, which are wide for READMEs. Lenient matching may accept some wrong claims, though only 1 of 100 checked found claims was wrong. Rule finding relied on a keyword filter and an LLM, which may miss rules phrased in other ways, such as a layer diagram. We did not measure recall against a full manual reading of every file, so our rules lean toward explicit, checkable ones. About half of the sentences that state a rule could not be checked, for example because they name parts that cannot be mapped to folders. Structured rules can also be wrong. We checked them in both directions: false breaks among all 51 rules flagged as broken, and hidden breaks in 80 of the 129 rules that held (Section~\ref{sec:step4}). Import resolution differs by language. Java and C\# packages do not always match folders, and custom bundler settings can hide imports, so some breaks are missed (5 of 40 checked intact rules). \emph{Mitigation:} we removed git-ignored paths, reviewed every structured rule, refined the codebook after a pilot, checked both flagged and unflagged cases, and report both automated and confirmed counts.

\paragraph{Internal validity.}
We detect agent-assisted commits only from visible traces. We combine message traces, agent accounts and pull-request traces, and human review confirmed 96 of 97 sampled agent labels. Still, agents that leave no trace stay hidden. Indeed, 3 of 53 sampled other commits did show a trace. So the comparison group contains some agent work, which makes the groups look more alike than they are \citep{robbes2026perils}. RQ3 is observational and supports no causal claim. Treating a path's last touching commit as its removal time can be wrong when a path is removed and re-added. \emph{Mitigation:} we followed renames, skipped the oldest commit of each clone, compared groups within projects, and dropped the few claims with impossible timelines.

\paragraph{Threats from the use of LLMs.}
Section~\ref{sec:llmdisc} discusses these threats in more detail. To summarize, LLM output can be wrong, can vary between runs and model versions, and two labellers using the same model, as ours did, can share errors. These risks affect the structured rules, the proposed review labels and the LLM-assisted analysis scripts. \emph{Mitigation:} No LLM output entered a result without human review. The first author reviewed every sampled item and set the final label, and a second reviewer verified the checks behind the main findings. A deterministic checker, not an LLM, decided whether each rule holds. The first author read and tested every script, and the review samples check script output against the repositories. We name the model and share the codebook and all raw outputs.

\paragraph{Reliability of human review.}
Human reviewers can also make mistakes, and judging what a rule's author meant requires interpretation. Our second reviewer saw the first author's labels, so their agreement may be higher than two fully independent labellings would give. Agreement was high ($\kappa$ 0.89 to 0.94), and the few disagreements did not change any final label. Checks of directly observable facts (paths, unflagged commits, agent labels) had only one human reviewer. \emph{Mitigation:} we used a written codebook, asked reviewers to judge from the evidence rather than the proposed label, gave every check that requires judging intent to a second reviewer, and report agreement for each. The labels and codebook are in the replication package.

\paragraph{External validity.}
We study open-source projects with project-level context files, over a short window (median 67 days per file). Results may differ for industrial code, for context files in subfolders, and over longer periods, when more claims may go stale. RQ3 rests on 7,480 commits in 96 projects but only 14 confirmed rule-breaking commits, so it is exploratory. \emph{Mitigation:} we used every engineered project with a context file in a large public dataset, covering all ten of its languages and five context-file formats. We compared against READMEs from the same projects.

\paragraph{Conclusion validity.}
With few rule-breaking commits, the test in RQ3 has limited power. A rough estimate suggests that detecting a difference of 1.8 times would need about seven times as many exposed commits. The absence of a significant difference is therefore not evidence that the groups behave the same. We treat the comparison as exploratory and report its direction, size and interval rather than a yes-or-no answer. \emph{Mitigation:} we used a test suited to rare events in grouped data, and we report its effect size with an interval.

\section{Future Work}
\label{sec:future}

We see three directions.

First, we plan to check richer architectural claims, such as responsibilities and data flow. Reflexion models \citep{murphy2001reflexion} combined with an LLM that maps descriptions to code are one way to do this. We also plan to study context files beyond the repository root.

Second, the discussion above raises a question about \emph{whose} understanding a context file records. We are running a study of shared understanding of architecture in student software teams that use AI assistants. There, we can ask team members how they see their system's architecture, and compare their views with each other, with the team's context file, and with the code over a project. This links context files to shared mental models \citep{cannonbowers1993smm,schelble2022letsthink} and to calls for shared knowledge spaces \citep{dearstyne2026darkmatter,akata2020hybrid}. Does writing a context file help a team build a shared model, or does it only record its author's view?

Third, controlled experiments could test what agents do when a rule is already broken in the code: follow the text, follow the code, or neither.

\section{Conclusion}
\label{sec:conclusion}

We checked the path claims and dependency rules in agent context files against the code and history of 3,346 open-source projects in ten languages.

Context files are good maps. Their path claims are mostly accurate, similar to READMEs, and are often updated in the same commit that moves the code.

They are weaker rulebooks. Almost one in five dependency rules was confirmed broken on the day it was written, usually by older code, and stayed broken. New rule breaks were rare. In an exploratory comparison, agent-assisted commits broke rules about 1.8 times as often as other commits, but with only 14 breaks the difference is not significant. This is neither evidence of a difference nor evidence that the two groups behave the same, and larger samples are needed to settle it.

We also showed how LLMs can support a mining study when people check every step. LLMs made the review of a large sample practical, but the decisions stayed with the human reviewers.

Keeping architectural claims true, and marking which ones are goals rather than facts, is a simple step toward developers and agents sharing the same picture of a system.

\section*{Statements and Declarations}

\paragraph{Funding.} Not applicable.

\paragraph{Competing interests.} The authors declare that they have no competing interests.

\paragraph{Ethics approval.} Not applicable. The study uses public repository data only.

\paragraph{Author contributions.} Bachan Ghimire designed the study, collected and analysed the data, carried out the human review, and wrote the manuscript. Daniela Damian supervised the study and reviewed and edited the manuscript.

\paragraph{Use of generative AI.} Section~\ref{sec:llmuse} describes how an LLM (Claude Opus~5.5) was used in the research. The authors reviewed all scripts and results and take full responsibility for the analysis.

\paragraph{Data availability.} The replication package contains the analysis pipeline, the extracted claims and dependency rules, the human labels and codebook, the prompt templates, all raw LLM outputs, and the analysis notebooks: \url{https://github.com/bachan48/github-context-mining/}. The input dataset is public \citep{galster2026dataset}.

\begin{acknowledgements}
We thank the colleague who served as the second reviewer for their careful verification of our labels.
\end{acknowledgements}

\bibliographystyle{spmpsci}
\bibliography{refs}

\end{document}